\documentclass[aps, prb, twocolumn]{revtex4-1}  
 
\usepackage{graphicx}  
\usepackage{bm}        
\usepackage{amssymb}   
\usepackage{amsmath}
\usepackage{epstopdf}

\newcommand{\ket}[1]{|#1\rangle}

\newcommand{\abs}[1]{ \left\lvert#1\right\rvert}

\begin{document}


\title{ Integrated optical multi-ion quantum logic}
\author{Karan K. Mehta$^*$}
\author{Chi Zhang}
\author{Maciej Malinowski} 
\author{Thanh-Long Nguyen}
\author{Martin Stadler}
\author{Jonathan P. Home}

\address{Department of Physics \& Institute for Quantum Electronics, ETH Z{\"u}rich, Switzerland}
\date{\today}

\maketitle

{\bf Practical and useful quantum information processing (QIP)  requires significant improvements with respect to current systems, both in error rates of basic operations and in scale. Individual trapped-ion\cite{leibfried2003quantum} qubits' fundamental qualities are promising for long-term systems \cite{haffner2008quantum}, but the optics involved in their precise control are a barrier to scaling \cite{monroe2013scaling}. Planar-fabricated optics integrated within ion trap devices can make such systems simultaneously more robust and parallelizable, as suggested by previous work with single ions \cite{mehta2016integrated}. Here we use scalable optics co-fabricated with a surface-electrode ion trap to achieve high-fidelity multi-ion quantum logic gates, often the limiting elements in building up the precise, large-scale entanglement essential to quantum computation. Light is efficiently delivered to a trap chip in a cryogenic environment via direct fibre coupling on multiple channels, eliminating the need for beam alignment into vacuum systems and cryostats and lending robustness to vibrations and beam pointing drifts. This allows us to perform ground-state laser cooling of ion motion, and to implement gates generating two-ion entangled states with fidelities $>99.3(2)\%$. This work demonstrates  hardware that reduces noise and drifts in sensitive quantum logic, and simultaneously offers a route to practical parallelization for high-fidelity quantum processors \cite{kielpinski2002architecture}. Similar devices may also find applications in neutral atom and ion-based quantum-sensing and timekeeping\cite{keller2019controlling}. }

Scaling quantum computing systems while continuing to improve operation fidelities over current systems is the central practical challenge in achieving general, large-scale quantum computation and simulation. Trapped-ion qubits'  long coherence times and high-fidelity gates \cite{harty2014high, ballance2016high, gaebler2016high} are likely to be valuable in managing overheads associated with quantum error correction \cite{fowler2012towards}, while their reliability and predictability as basic quantum systems may assist in preserving performance as complexity grows. However, the bulk optics arranged over meters-long free-space paths typically used for trapped-ion control are a major source of drift and noise, and are furthermore challenging to parallelize, particularly in 2D trap geometries \cite{amini2010toward}. 

Previous work has demonstrated planar-fabricated nanophotonic waveguides integrated with surface-electrode ion traps that offer advantages over free-space approaches in beam pointing and phase stability,  power usage, addressability, and parallelizability\cite{mehta2016integrated}. However, the free-space input coupling and low efficiencies in the early work (33 dB loss from input to ion) compromised the potential advantages in stability and limited the utility of the delivered light for spin-motion coupling, which is essential for both ground-state cooling of ion motion and multi-qubit quantum logic. 

In this Article, we use integrated optics to drive multi-ion entangling quantum logic with fidelities competitive with those in state-of-the-art experiments across qubit platforms. Our experiments leverage efficient interfacing of multiple input fibres to integrated waveguides in a cryogenic ion trap chip ($2.4$ dB loss at optical wavelength $\lambda = 729$ nm after cooling to 7K) via direct attachment. We characterize limiting noise sources, finding significant room for future improvements. The scalability and precision simultaneously afforded with this approach is promising for pursuit of large-scale computation and simulation, and we anticipate similar techniques will find broad application across a range of atom-based technologies. 

\begin{figure}[b]
\centerline{\includegraphics[width=.45\textwidth]{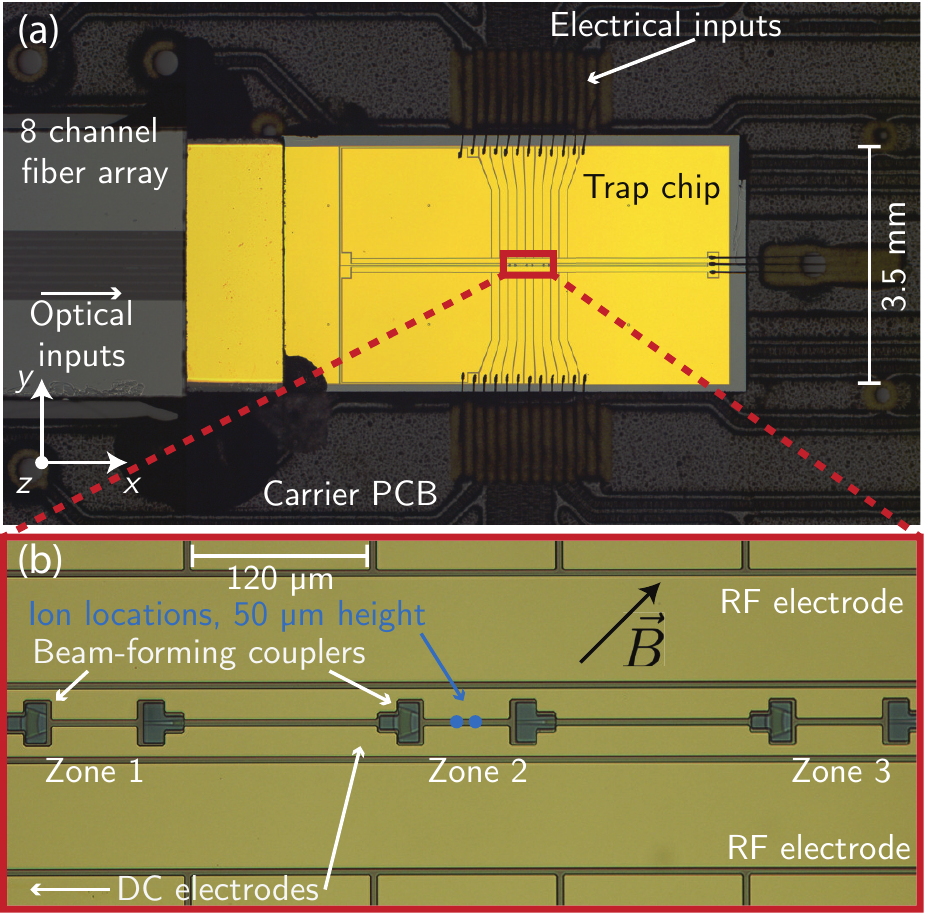}}
\vspace{0 cm}
\caption{\label{fig:2qdev} \textbf{Device overview}. (a) Optical micrograph of the assembled ion trap device with eight-channel fibre array attached; optical fibres extend from the device to a vacuum feedthrough, with optical inputs via standard connectors outside the vacuum chamber. (b) Higher magnification view near the trap zones, showing electrode openings for light from waveguide couplers at the three trap zones (zone 3 was used in the experiments below).}
\end{figure}

\begin{figure*}[t]
\centerline{\includegraphics[width=\textwidth]{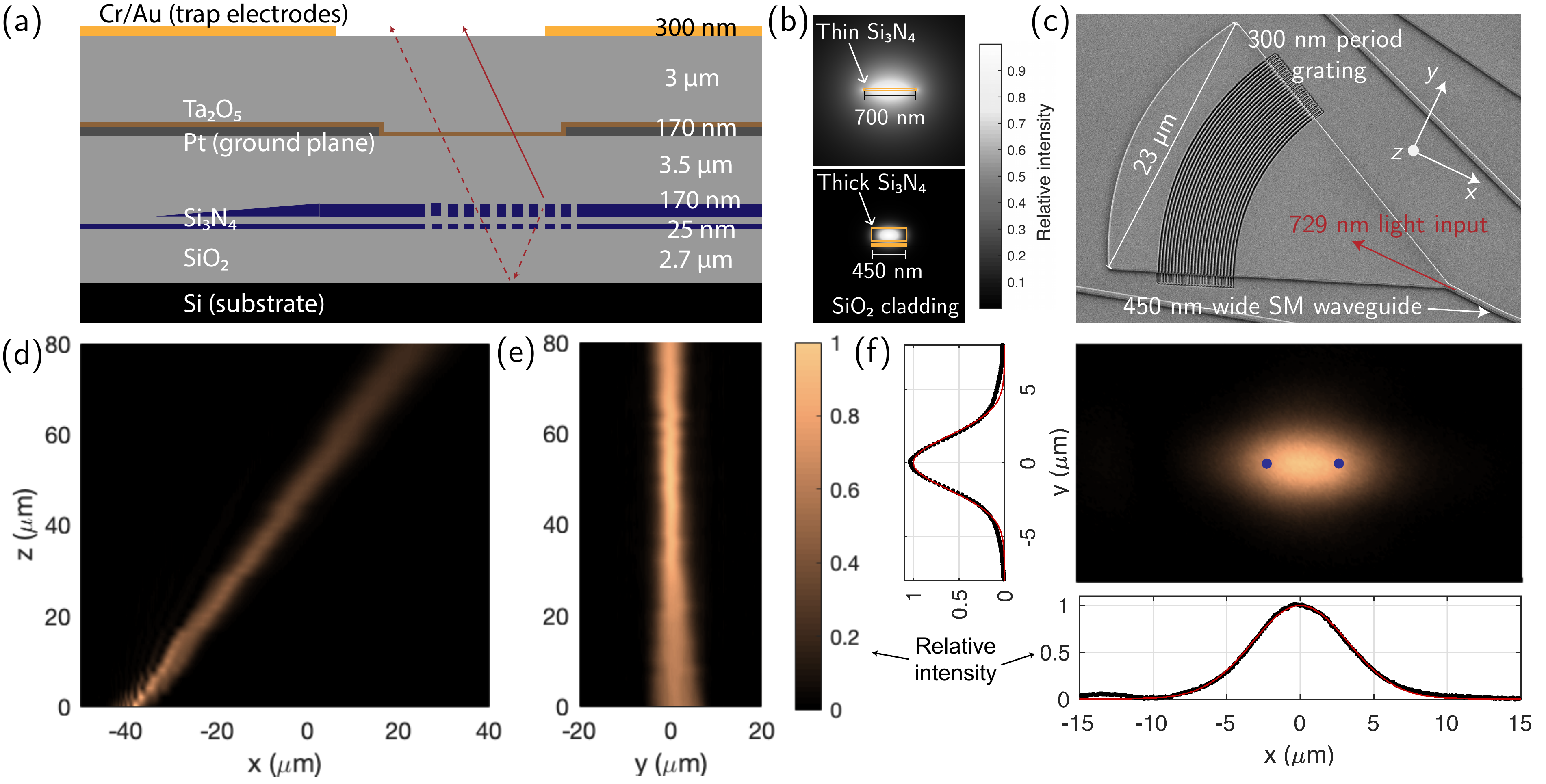}}
\vspace{0 cm}
\caption{\label{fig:gratings} \textbf{Layer stackup and optical design}.  (a) Layer stackup in trap fabrication. The top gold layer is patterned to form trap electrodes, to which RF and DC potentials are applied to confine ions 50 $\mu$m above the surface. The lower platinum serves as a ground plane, and waveguides/gratings are formed in the SiN. The Si substrate is additionally used as a bottom reflector to increase the grating efficiency \cite{mehta2017precise}. (b) Simulated intensity profiles of quasi-TE modes ($\vec{E}$ polarized predominantly along the horizontal) of the two waveguide cross-sections, with core areas outlined in gold. (c) Scanning-electron-microscope image of a fabricated grating, showing photolithographically defined waveguides/tapers, and e-beam written grating lines with 300 nm period. (d) and (e) 729 nm-beam emission profiles in the $xz$ and $yz$ planes, obtained from image stacks of radiated field at various heights $z$ above the trap electrode layer. (f) Intensity pattern at the the trap height $z=50$ $\mu$m; radiation is polarized predominantly along $\mathbf{\hat y}$, and Gaussian fits indicate $1/e^2$ intensity radii in the $xy$ plane of $w_x = $ 6.5 $\mu$m and $w_y = 3.7$ $\mu$m. Blue dots label trapped ion positions in the radiated beam and confining potential applied in experiments below.}
\end{figure*}

We designed trap devices with integrated photonics \cite{mehta2019towards} for fabrication in an openly accessible commercial foundry  \cite{worhoff2015triplex}. An optical micrograph of the device studied in this work is shown in Fig.~\ref{fig:2qdev}. A surface-electrode Paul trap \cite{chiaverini2005surface} is formed in the top gold layer of a microfabricated chip, and designed to confine $^{40}$Ca$^+$ ions 50 $\mu$m above the surface at the three zones labeled in Fig.~\ref{fig:2qdev}b. The trapping potential is formed by radiofrequency (RF) and DC potentials applied to the electrodes. Light controlling the ions is input to the device via an attached multi-channel fibre V-groove array. 

Fig.~\ref{fig:gratings}a shows the layers comprising the device. A platinum layer forms a ground plane beneath the trap electrodes; this is intended to shield the silicon substrate from the trap RF fields, avoiding modulation of the trap impedance (and hence voltage supplied to the trap and confining potential) during operation that may result from photoconductivity in the silicon\cite{mehta2014ion}. This ground layer also assists in shielding ions from mobile charge carriers in the substrate. Between this ground plane layer and the substrate, thin-film silicon nitride (SiN) is patterned to form  waveguides and gratings, with structures implemented for wavelengths used for coherent qubit control (729 nm) and repumping (854 and 866 nm wavelengths) \cite{schindler2013quantum}. 

A thin 25 nm SiN layer supports a weakly-confined mode with mode area matched to a standard single-mode fibre at these wavelengths (Fig.~\ref{fig:gratings}b); these waveguides extend to the chip edge to interface to input/output optical fibres. To route light on chip, power in this mode is coupled to a highly-confined mode in a thicker core (consisting of an additional 170 nm-thick layer above) via a ${\sim}1$ mm adiabatic taper \cite{worhoff2015triplex}. These structures allow coupling from multiple attached fibres to on-chip waveguides, with 1.4 dB measured insertion loss at room temperature. Fibre feedthroughs on the outer vacuum chamber then allow precise optical delivery to ions in the cryostat by simply connecting optical fibres outside the vacuum chamber. This beam delivery also suppresses noise from beam-pointing fluctuations and cryostat vibrations (see Methods). 

Input light is routed to the trap zones shown in Fig.~\ref{fig:2qdev}b, where a waveguide taper expands the lateral mode diameter and a series of curved grating lines serve to emit light into a designed beam. Openings in the trap electrodes transmit the light and are included symmetrically around each trap zone; simulations indicate negligible impact of these windows on the RF pseudopotential in this design. 

Gratings (Fig.~\ref{fig:gratings}c) for 729 nm light are designed to uniformly illuminate two ions oriented along the trap axis $\mathbf{\hat x}$, and hence focus along $\mathbf{\hat y}$ and emit a roughly collimated beam along $\mathbf{\hat x}$. Grating emission is characterized by scanning the focal plane of a high-NA microscope up from the chip surface and capturing a stack of images of the emitted intensity profile at various heights \cite{mehta2016integrated, mehta2017precise} (see Methods). Cross-sectional profiles, and the beam profile at the ion trap height of $z=50$ $\mu$m measured with this method, are shown in Fig.~\ref{fig:gratings}c-e. 

After the trap die is wirebonded to a carrier princted circuit board (PCB) and the fibre array is attached, the resulting assembly is mounted in a cryogenic vacuum apparatus and cooled to ${\sim}7$ K for ion trap experiments. The fibre array is attached in a fashion that allows us to maintain low fibre-chip coupling losses on each channel despite the temperature drop from 300 K (see Methods).   

Light emitted by the 729 nm couplers drives all transitions between $4S_\frac12$ and $3D_\frac52$  with $\abs{\Delta m_J} \le 2$ because the light emitted by the grating couplers contains a mixture of $\mathbf{\hat \sigma_+}$, $ \mathbf{\hat \sigma_-}$, and $\mathbf{\hat \pi}$ polarizations relative to the quantization axis defined by the magnetic field of ${\sim} 6$ G along $\mathbf{\hat{x}} + \mathbf{\hat{y}}$ (indicated in Fig.~\ref{fig:2qdev}b). For the experiments below, we choose qubit states $\ket{{\downarrow}} = \ket{4S_\frac12,m_J = -\frac12}$ and $\ket{{\uparrow}} = \ket{3D_\frac52,m_J = -\frac12}$, owing to the relatively low differential Zeeman shift (and hence minimal sensitivity to magnetic field fluctuations) of the corresponding transition (0.56 MHz/G). 

In our experiments, two ions are confined in a potential well with axial center-of-mass (COM) mode frequency $\omega_\mathrm{COM} =2\pi\times1.2$ MHz (along $\mathbf{\hat x}$), with a resulting ion spacing of 5 $\mu$m. Radial mode frequencies are between 3.7 and 5.7 MHz. In each experimental shot, all motional modes are Doppler cooled via conventionally-delivered free-space beams, and axial motion is further cooled with electromagnetically-induced-transparency (EIT) cooling \cite{roos2000experimental}. This leaves axial modes with mean phonon numbers $\bar n \approx 0.5$, and radial modes with $\bar n \approx 5-15$. The qubit is prepared in the $\ket{{\downarrow}}$ state via frequency-selective optical pumping on the $ 4S_\frac12 \leftrightarrow 3D_\frac52$ transition with infidelities $<10^{-4}$ (see Methods). 

\begin{figure}[t]
\centerline{\includegraphics[width=0.5\textwidth]{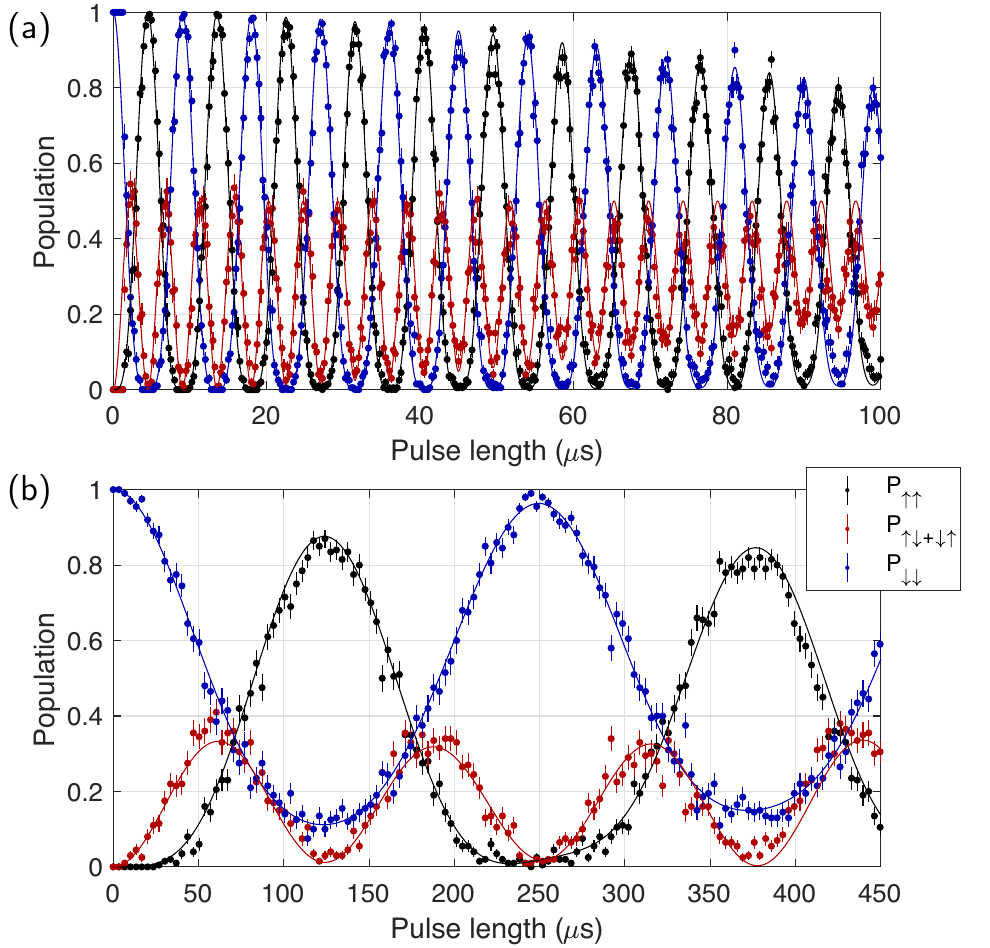}}
\vspace{0 cm}
\caption{\label{fig:2ion} \textbf{Two-ion manipulation and ground-state cooling}. (a) Coherent manipulation of two ions with laser frequency $\omega$ resonant with the carrier transition at $\omega_0$. Fits to theory (solid lines) indicate Rabi frequency imbalance between the two ions $<1\times10^{-2}$. (b) Population evolutions upon application of a pulse resonant with the blue sideband ($\omega = \omega_0 + \omega_\mathrm{STR})$ on the stretch mode, after sideband cooling with light coupled via the integrated couplers. Lines correspond to theoretical evolutions with $\bar{n}_\mathrm{STR} = 0.05$. In both (a) and (b), each point represents an average over 200 experimental shots, with error-bars indicating 1$\sigma$ standard errors from projection noise.}
\end{figure}

Coherent single-qubit rotations on two ions resulting from a 729 nm pulse resonant with the qubit transition  following state preparation and cooling are shown in Fig.~\ref{fig:2ion}a. Observed Rabi frequencies are consistent with expectation \cite{james1998quantum}, given this beam profile, power, and total loss (6.4 dB input-to-ion, see Methods). DC potentials are applied to translate ions along $\mathbf{\hat x}$ to positions where the beam intensity on each ion is equal (Fig.~\ref{fig:gratings}f). The observed Rabi oscillations allow us to bound the Rabi frequency imbalance ($\abs{1 - \Omega_1/\Omega_2}$) seen by the two ions to $<1\times10^{-2}$. This balancing is sensitive to ion positioning at the level of ${\sim}100$ nm. The exposed dielectric on the trap surface owing to the grating windows and gaps between electrodes may become charged and generate stray fields that could shift ion positions, but we nevertheless observe that this positioning is stable over hours of experimenting with two-ion chains. The decay in Rabi oscillation amplitude is consistent with that expected given the thermal occupancy of the Doppler-cooled radial modes \cite{wineland1998experimental}. 

Ramsey measurements on single ions were performed to assess qubit coherence. With the integrated 729 nm beam path, we observe $T_2^*$ contrast decay times (see Extended Data Fig.~\ref{fig:ramsey} and Methods) significantly longer than observed using the same light addressing the ion through a free-space path. These decays are dominated by phase noise in our 729 nm light. We attribute the enhancement in coherence to the fact that in the free-space configuration, cryostat vibrations relative to the optical table result in laser phase fluctuation at the ion, a source of noise which is suppressed for the integrated beam path due to the common motion of the grating and trap.  

\begin{figure*}[t]
\centerline{\includegraphics[width=\textwidth]{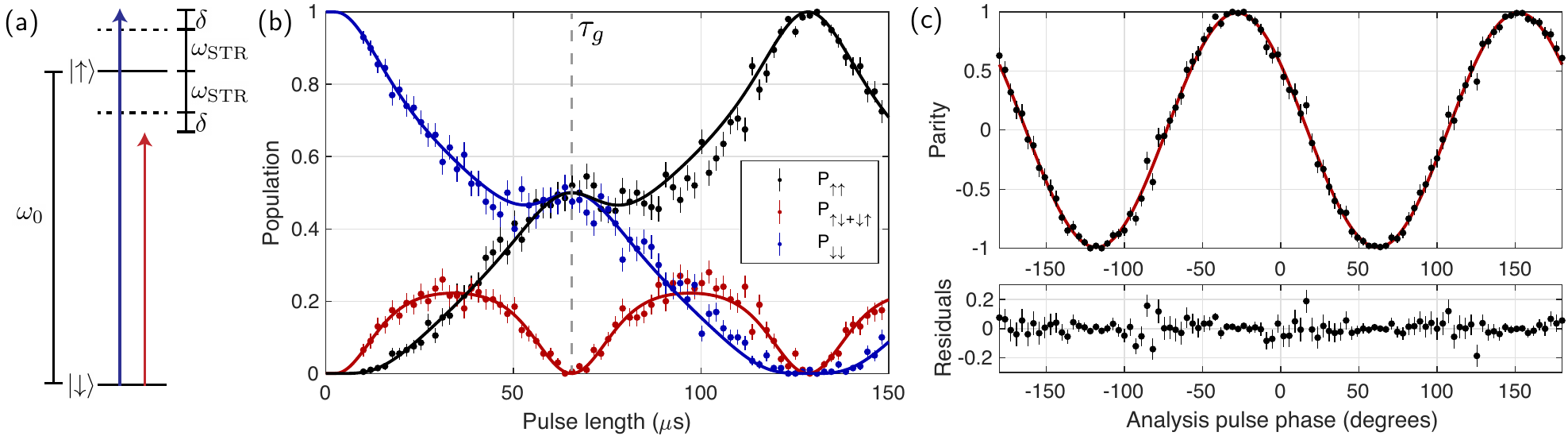}}
\vspace{0 cm}
\caption{\label{fig:gate} \textbf{Integrated implementation of a two-ion quantum logic gate.} (a) Optical frequencies applied to drive a M{\o}lmer-S{\o}rensen gate; in our experiments $\omega_0 = 2\pi\times 411$ THz (729 nm), $\omega_\mathrm{STR} = 2\pi\times2.0$ MHz, and the detuning from motional sidebands is $\delta = 2\pi\times15$ kHz. (b) Population evolutions as a function of pulse duration; solid lines show ideal gate evolutions from theory, with gate time as the only free parameter. (c) Parity oscillations observed upon scanning the phase of an ``analysis" $\pi/2$ pulse after the gate time $\tau_g = 66$ $\mu$s. The fit contrast (red line) of 0.992(2), together with the even populations measured at the gate time 0.994(1), indicate a Bell state generated with fidelity 0.993(2). In both (b) and (c), each point represents an average over 200 experimental shots, with error bars indicating 1$\sigma$ standard errors from projection noise. }
\end{figure*}

Multi-ion quantum logic gates are commonly mediated by shared motional modes; high-fidelity implementations typically require near-ground-state cooling, particularly of the modes used to enact the gates. We implement gates using axial motion, and choose the ``stretch" mode in which ions oscillate out of phase (at $\omega_\mathrm{STR} = \sqrt{3}\omega_\mathrm{COM}$), owing to its relative insensitivity to electric-field noise that heats ion motion \cite{brownnutt2015ion}. We use the light delivered by the integrated couplers to implement pulsed sideband cooling \cite{wineland1998experimental}. Blue sideband flopping following ground-state cooling (Fig.~\ref{fig:2ion}b) indicates a mean phonon number $\bar{n}_\mathrm{STR} \approx 0.05$. 

We proceed to implement entangling operations mediated by the ground-state-cooled stretch mode. M{\o}lmer-S{\o}rensen gates \cite{sorensen2000entanglement} apply the unitary $\hat{U}_\mathrm{MS}(\phi) = \exp \left[-i \pi (\sum_k \hat{\sigma}_{\phi,k})^2/8 \right]$, with $\hat{\sigma}_{\phi,k} = \cos(\phi)\hat\sigma_{x,k} + \sin(\phi)\hat\sigma_{y,k} $  representing Pauli operators on qubit $k$ (with $\ket{{\uparrow}}$ and $\ket{{\downarrow}}$ the eigenstates of $\hat\sigma_z$). The spin phase $\phi$ is set by laser phases. We implement $\hat{U}_\mathrm{MS}$ by applying laser light at two frequencies detuned from the stretch mode sidebands of the 729 nm carrier transition at $\omega_\pm = \omega_0 \pm (\omega_\mathrm{STR} + \delta)$ (see Fig.~\ref{fig:gate}). Starting from $\ket{{\downarrow \downarrow}}$, the ideal implementation generates the Bell state $\frac{1}{\sqrt{2}}\left(\ket{{\downarrow \downarrow}} - i\ket{{\uparrow \uparrow}}   \right)$ after a pulse time $\tau_g = 2\pi/\delta$. Measured population evolutions as a function of pulse duration are shown in Fig.~\ref{fig:gate}b, together with theoretical predictions for an ideal gate. To measure the coherence of the state produced at $\tau_g = 66$ $\mu$s, we apply identical single qubit rotations $\hat{R}_k(\theta, \phi_a) = \exp(-i\theta\hat\sigma_{\phi_{a},k}/2)$ to both ions ($k=1,2$) with $\theta = \pi/2$ after the gate, with variable phase $\phi_a$. These ``analysis" pulses map the Bell state to states whose parity varies sinusoidally with $\phi_a$; here, parity is defined as $\mathcal{P} = P_{\uparrow\uparrow} + P_{\downarrow\downarrow} - P_{\uparrow\downarrow} - P_{\downarrow\uparrow}$, where $P$s are the populations measured in the four possible two-ion outcomes. The contrast of the parity oscillations as a function of $\phi_a$ is a direct measure of the coherence of the Bell state generated at $\tau_g$ \cite{leibfried2003experimental}. A maximum-likelihood fit to the data (Fig.~\ref{fig:gate}c) indicates a contrast of  99.2(2)\%, together with the populations measured at the gate time (99.4(1)\%) indicating a total Bell-state fidelity of 99.3(2)\%, not correcting for state preparation and measurement error. 

Table~\ref{tab:errors} summarizes contributing error sources, which are detailed in the Methods. Leading infidelities arise from motional heating and mode frequency drifts, and laser frequency noise. Measurements of single-ion heating in our device indicate $E$-field noise ${\sim}100\times$ higher than in other cryogenic planar traps with similar ion-electrode distance \cite{sedlacek2018distance}, suggesting significant potential for reduction of this noise. Various technical improvements could reduce motional and laser frequency drifts in our apparatus; though we have implemented the simplest-possible single-loop gate here, multi-loop and composite-pulse gates can further suppress errors resulting from these drifts \cite{milne2020phase}. The only error source fundamental to this choice of qubit is spontaneous emission  from the $D_\frac52$ level, which given its 1.1s lifetime contributes an error of $3\times 10^{-5}$ for the initial state and 66 $\mu$s gate time employed here (this error scales with $\tau_g$). 

\begin{table}[b]
\centering
\begin{tabular} {| c | c |} 
\hline
\textbf{\hspace{40pt} Error source \hspace{40pt}} & \textbf{Infidelity ($\times 10^{-3}$)} \\
\hline
Motional mode heating & $2(1)$ \\
\hline
Motional frequency drifts & $1 $ \\
\hline
Laser frequency noise & $1$ \\
\hline
Two-ion readout error & $0.9$ \\
\hline
Kerr cross-coupling & $0.4 $ \\
\hline
Spectator mode occupancies & $0.3 $ \\
\hline
Spontaneous emission & $0.03$ \\
\hline
\hline
\textbf{Total} &  $\mathbf{{\sim}6\times 10^{-3}}$\\
\hline
\end{tabular}
\caption{\label{tab:errors} \textbf{Error budget.} Error sources and estimated contributions to infidelity of the Bell states generated in this work. }
\end{table}

The infidelities achieved in this work are within an order of magnitude of the lowest demonstrated \cite{gaebler2016high, ballance2016high}. While microwave-based schemes are promising for achieving the highest possible fidelities \cite{srinivas2019trapped}, we note that the mW-level optical powers in our work are 2-3 orders of magnitude lower than the powers required in microwave gate implementations in planar traps to date \cite{harty2016high, zarantonello2019robust}, in which gate times were additionally an order of magnitude longer. Power requirements are likely to be critical in larger-scale systems where dissipation is a concern. The short wavelengths of optical radiation also allows simple addressing of particular ions or trap zones \cite{debnath2016demonstration, mehta2017precise} without active cancellation. When inputting light to the coupler addressing one zone in the present device, crosstalk at neighboring zones manifests at the $-60$ dB level in relative intensity (see Methods). However, spontaneous emission errors and requirements on long-term laser phase stability may eventually prove problematic in deep quantum circuits; these issues can be alleviated in ions like $^{40}$Ca$^+$ by utilizing both the Zeeman and optical qubit transitions (see Methods). 

This work motivates developments along multiple fronts. Wavelengths guided in the current chip should allow laser cooling and readout using IR and high-intensity visible light \cite{lindenfelser2017cooling}, while materials and waveguides for blue/UV light \cite{west2019low} allow photoionization, laser cooling, and readout at low intensity (recent contemporary work \cite{niffenegger2020integrated} demonstrates these functionalities for single-qubit operations, using PECVD SiN waveguides). This may enable practical, parallel operation of multiple zones with fully integrated beam delivery, in a fashion also extensible to dual-species systems and 2D trap architectures; transport between zones would be a natural step towards large-scale computations \cite{kielpinski2002architecture, kaufmann2017scalable}. The efficient cryogenic fibre-to-chip coupling shown here, together with stable tight focusing \cite{mehta2017precise}, will enable practical, power-efficient fast optical gates \cite{schafer2018fast}, while combination of optics for incoherent operations with microwave electrodes could facilitate scaling approaches using microwave coherent manipulation \cite{harty2016high, zarantonello2019robust, srinivas2019trapped}. Finally, the precise ion positioning demonstrated here indicates the practicality of schemes utilizing integrated optics' ability to generate spatially structured fields \cite{mehta2019towards}.  

Beyond their broad relevance in trapped-ion QIP, these approaches may also benefit atomic clocks, whether in portable systems or in precise clocks utilizing multiple trap zones \cite{keller2019controlling}, as well as neutral atom quantum systems \cite{saffman2010quantum}.

\appendix*
\section{Methods Summary}

\subsection{Device design and fabrication}
Devices were designed similarly to those in previous work\cite{mehta2017precise}. The grating designs exploit the reflection from the silicon substrate below to increase grating strength; a compromise between grating length and efficiency resulted in a designed radiation efficiency of 50\% in these devices (defined as upwards-radiated power as a fraction of input power). Emission into a single grating order is ensured by choosing a sufficiently large grating wavenumber $\beta_g = \frac{2\pi}{\Lambda}$ (with $\Lambda$ the grating period). The curvature of the grating lines is calculated to induce focusing along $\mathbf{\hat{y}}$ in Fig.~\ref{fig:2qdev}b, with designed focusing limited to ${\sim}3$ $\mu$m waists to relax sensitivity to misalignments in fabrication. $\Lambda$ is constant over the grating length in these devices to generate an approximately collimated beam along the trap axis. 

\begin{figure}[b]
\centerline{\includegraphics[width=0.5\textwidth]{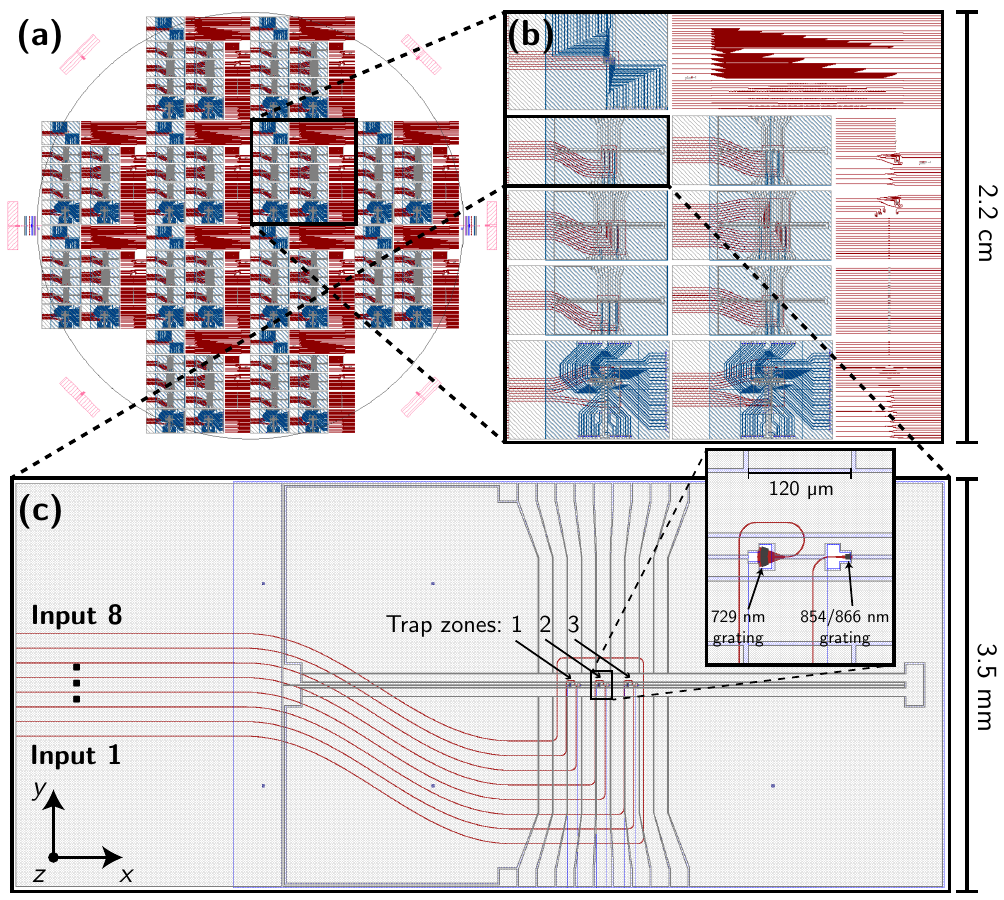}}
\vspace{0 cm}
\caption{\label{fig:wafer} \textbf{Design layout}. (a) Mask images for device fabrication across a 4-inch wafer. (b) Individual $2.2\times2.2$ cm$^2$ reticle, showing trap designs as well as independent optics test structures. (c) Trap design used in ion experiments presented here. In all images, SiN features are shown in red, the top trap electrode layer in gray, and the ground plane in blue. The 8 waveguides coupled to the fibre array are labeled at the left, with inputs 1 and 8 forming a loop structure used to align the fibre V-groove array. }
\end{figure}

Mode overlap simulations predict a 1 dB coupling loss between an optical mode with 5.4 $\mu$m mode-field diameter and the waveguide mode of the 25 nm-thick SiN waveguides. Waveguides formed of the thin SiN core in the fibre coupling-regions are strongly polarizing in our devices, in that owing to the metal and silicon a few $\mu$m from the core, only the mode with E-field polarized predominantly along the horizontal in Fig.~\ref{fig:gratings}b propagates in these regions with low loss. Polarization is maintained in the high-confined waveguides due to the significant index mismatch between the quasi-TE and quasi-TM modes in these regions, typical for integrated waveguides with non-square aspect ratios. 

Grating designs were verified with 3D finite-difference-time-domain (FDTD) simulations (Lumerical FDTD solutions), and 3D simulations of the ion trap potentials were performed in COMSOL; designs were drawn in Cadence Virtuoso.

Various trap and optical designs were drawn on a $2.2\times2.2$ cm$^2$ reticle repeated across a 4-inch wafer (Extended Data Fig.~\ref{fig:wafer}). A design drawing of the trap die used in this work is shown in Extended Data Fig.~\ref{fig:wafer}c, together with a magnified view near one of the trap zones with 729 nm and 854/866 nm waveguides and gratings. The fabrication process used here allowed relative alignment between different layers within approximately $\pm$2 $\mu$m, and the e-beam grating lines were aligned within about 300 nm to the photolithographically-defined waveguides. To account for possible misalignments, grating lines in zones 1 and 3 were intentionally offset along $y$ by $\pm 300$ nm. All three zones were characterized in optical measurements; the 300 nm offset between grating and waveguide features results in a ${\sim}2.5$ $\mu$m beam shift along $y$ at the ion height between zones, in accordance with simulations. 

Fabrication was performed by LioniX International \cite{worhoff2015triplex}. Devices were fabricated on silicon wafers with 5-10 Ohm-cm resistivity; the bottom 2.7 $\mu$m oxide layer was formed via thermal oxidation, with the waveguides formed in LPCVD silicon nitride layers. Following combined stepper photolithography and e-beam patterning of the waveguide/grating features, LPCVD SiO$_2$ cladding was deposited. The platinum layer was then deposited after planarization, and patterned via contact photolithography ion beam etching. A thin (90 nm) Ta$_2$O$_5$ layer serves as an adhesion layer above the Pt. The upper PECVD SiO$_2$ isolation was patterned to allow vias between the two metal layers, after which the upper gold was sputtered and patterned via contact photolithography and a liftoff procedure to form the trap electrodes.  

Diced wafers were delivered to ETH, where die were characterized for optical performance, or else cleaned, packaged and fibre-attached for ion trap experiments.

\subsection{Assembly and fibre attachment}
A standard eight-channel fibre V-groove array populated with Nufern S630-HP fibres spaced at 127 $\mu$m pitch was purchased from OZ Optics. Eight waveguides extend to the edge of the chip die to interface to the array, with the outer two forming a simple loop structure used to align the array by maximizing loop transmission. Standard non-polarization-maintaining fibre is used in this work, and in-line fibre polarizers control the input polarization. 

\begin{figure}[t]
\centerline{\includegraphics[width=0.45\textwidth]{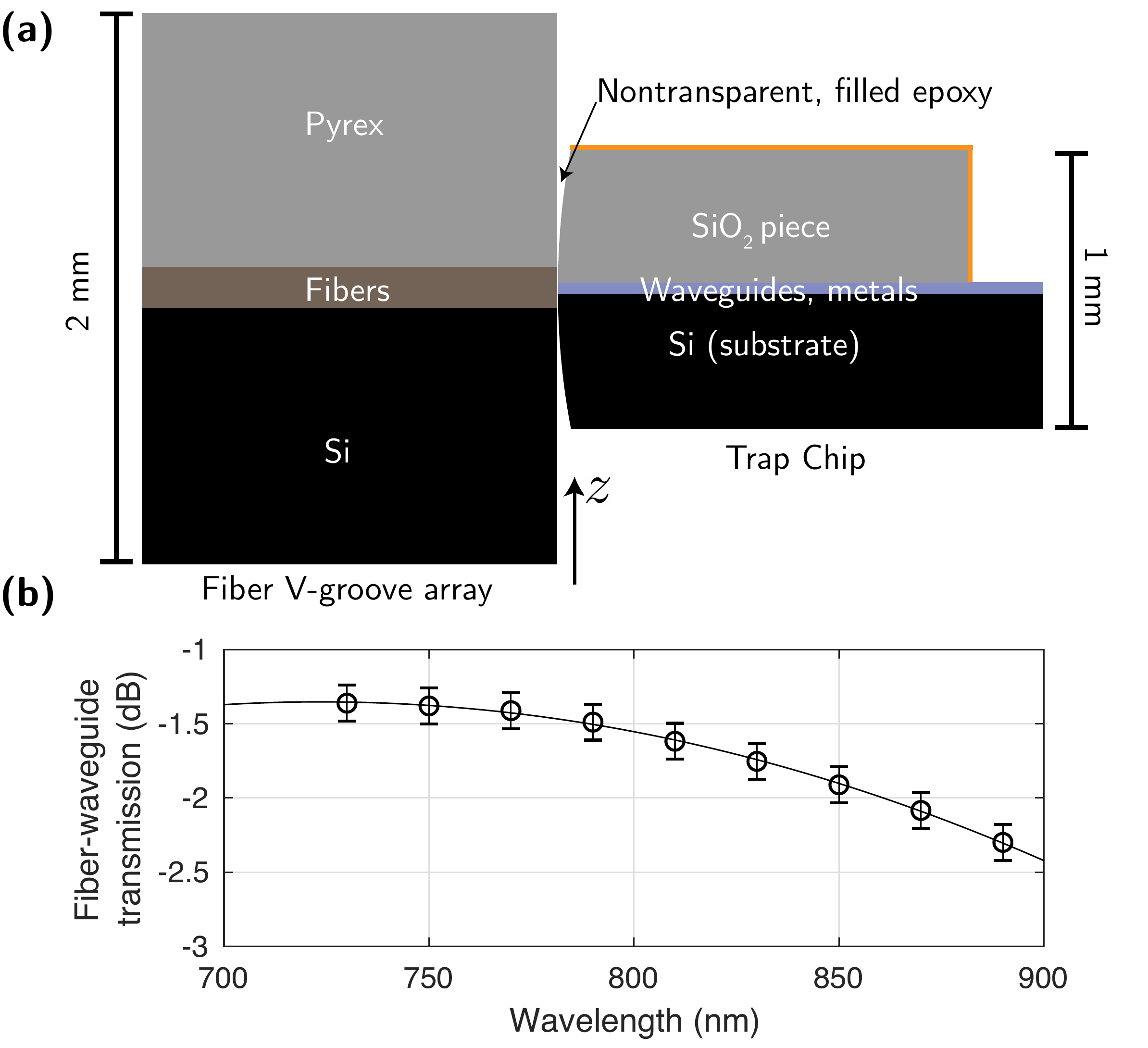}}
\vspace{0 cm}
\caption{\label{fig:fibattach} \textbf{Fibre attachment}. Fibre attach process schematic and measured single pass fibre-waveguide coupling losses inferred from a loop-back structure on-chip; solid line is a guide to the eye. }
\end{figure}

Individual trap die were removed from the wafer, and a $1.5 \times 3.5$ mm$^2$ SiO$_2$ piece of 500 $\mu$m thickness was epoxied (EPO-TEK silver epoxy, model H21D) to the top surface of the trap chip at the fibre-coupling edge (Extended Data Fig.~\ref{fig:fibattach}, and visible near the fibre array in Fig.~\ref{fig:2qdev}a). This piece was previously coated in 300 nm Au via electron-beam evaporation on faces exposed to the ion, to minimize possible stray fields. Subsequently the die was coated in a protective photoresist layer and mounted in a custom-machined holder for polishing on standard fibre polishing paper. Polishing reduces roughness and associated loss at the coupling interface, and introduces convex curvature to the facet, important for the fibre attachment as described below. The silicon substrate was scribed with a diamond scribe to break through the native oxide; after dissolving the protective photoresist and attaching the die to the carrier PCB, a drop of silver epoxy was applied both to contact the silicon substrate to ground on the carrier PCB, as well as to ensure grounding of the metal on the SiO$_2$ piece near the fibre array. The trap electrodes were then wirebonded to the contacts on the carrier PCB. 

To attach the fibre array, the facet-polished and wirebonded sample was mounted on a temperature-controlled chuck fixed to a 3-axis tip/tilt/rotation stage and raised to 95$^\circ$C. The fibre array, held in a custom machined stainless-steel mount in a 3-axis translation stage, was then aligned and pressed against the chip in such a fashion that static friction between the two interfaces allowed coupling to be passively maintained for hours (see Extended Data Fig.~\ref{fig:fibattach}a for a schematic of this interface). Low-temperature curing, non-transparent, filled epoxy (EPO-TEK T7109-19) was then dropped at the two edges and allowed to wick into the gap; this epoxy was chosen as its flexibility should result in robustness to temperature changes. The convex curvature of the trap die at this interface ensures that the epoxy does not interact with the optical mode, allowing choice of mechanically suitable epoxies without restriction to transparent variants; minimizing exposure of the optical mode to the epoxy is also important in avoiding possible photo-effects which can be problematic even with transparent adhesives exposed to visible wavelengths. The total contact area used for the attachment is about $1\times3.5$ mm$^2$. After a few hours of curing with the sample stage at $95^\circ$C, the holder for the fibre array was loosened to allow the fibre array to move with the trap chip, the sample stage was cooled to room temperature, and no drop in transmission was observed. We have performed three attachments in this fashion to date, and found all robust to temperature drops. We note that in this work the epoxy was applied manually -- more precise automatic dispensing should allow more symmetric application, which may increase robustness to temperature changes further still. 

The transmission of a single fibre-waveguide interface was measured at various wavelengths within the bandwidth of a CW Ti:sapphire laser, measuring the transmission from input fibre 1 to fibre 8 and subtracting the measured waveguide loss. The broadband nature of this coupling is shown in the measurements in Extended Data Fig.~\ref{fig:fibattach}b. Upon applying current to a heater on the carrier PCB near the chip and then cooling the assembly to 7K, we observed a small increase in loss from 1.4 dB to 2.4 dB at 729 nm, but saw no further changes after two additional temperature cycles between room temperature and 7K. These coupling losses compare favorably to previous published work on cryogenic fibre attachment \cite{mckenna2019cryogenic}. This method furthermore allows multiple channels aligned in parallel, and at shorter wavelengths with more demanding alignment tolerances. 

\subsection{Waveguide/grating characterization and optical losses}
Structures consisting of waveguides of varying length on chip were included on the reticle design to measure waveguide losses in the high-confinement mode used for routing on chip. These were measured to be 2 dB/cm at 729 nm, and verified with measurements of quality factors of ring resonators also included in the design. Wavelength-dependent losses were measured from 729 to 920 nm using a tunable CW Ti:sapphire source; we observe a reduction in loss as a function of increasing wavelength consistent with scattering from sidewall roughness being the dominant contribution. We therefore expect improved lithography to reduce this propagation loss. 

Grating emission was profiled (Fig.~\ref{fig:gratings}c-e) using a microscope mounted on a vertical translation stage to image the emission at various heights above the chip\cite{mehta2017precise} onto a scientific CCD camera (Lumenera Infinity 3S-1UR), through a 0.95-NA objective (Olympus MPLANAPO50x). An RF modulation input was applied to the current drive of a 730 nm diode laser to reduce the coherence length to avoid reflection artifacts in the imaging system. The measured height of the beam focus above the top metal layer agrees with FDTD simulation to within the measurement accuracy of about 2 $\mu$m, and the measured emission angle in the $xz$-plane (Fig.~\ref{fig:gratings}d) matches design to within about 1$^\circ$. These measurements are conducted at room temperature; considering the thermo-optic coefficients of SiO$_2$ and SiN \cite{elshaari2016thermo}, we expect a change in emission angle upon cooling to 7K of roughly $0.2^\circ$, negligible in our geometry. 

With 1.5 mW of 729 nm light input to the fibre outside the cryostat, we observe a 2.6 $\mu$s $\pi$-time, which is within 25\% of the 2.0 $\mu$s predicted from a first-principles calculation \cite{james1998quantum} accounting for the measured beam profile, and the total expected loss of 6.4 dB, arising from: 2.4 dB fibre-chip coupling, 1 dB of waveguide loss over the 5 mm path length, and 3 dB grating emission loss. In the $\pi$-time calculation we assume the ion sits at the maximum of the 3.7 $\mu$m beam waist along $\mathbf{\hat{y}}$, resulting in a lower-bound given the $\mu$m-scale misalignments in these devices. Similar $\pi$-times are observed in the two-ion Rabi oscillations in Fig.~\ref{fig:2ion}a; we note that in this data, the $P_{\uparrow\downarrow + \downarrow\uparrow}$ points are particularly helpful in bounding the Rabi frequency imbalance between the two ions, as a small imbalance results in these populations rising significantly above 0.5 for pulses much longer than the $\pi$-time. 

In testing power handling of these waveguides, a maximum of 300 mW was coupled into the single-mode waveguides at $\lambda=729$ nm; we did not observe damage to the waveguides from these intensities. Future work will investigate power handling limits and self-phase modulation in such waveguides at visible wavelengths, relevant to large-scale architectures with power for multiple zones input to a single bus waveguide. 

\subsection{Laser/cryogenic apparatus, and trap operation}

Diode lasers supply the various wavelengths used in the ion experiments. The 729 nm light used for coherent qubit control is stabilized to a high-finesse reference cavity resulting in a linewidth of order 100 Hz. Light transmitted through the cavity injection-locks a secondary diode, whose output passes through a tapered amplifier (TA) to a free-space double-pass acousto-optic modulator (AOM) setup for switching and frequency tuning, and subsequently a single-pass fibre-coupled AOM for pulse-shaping. Two tones are applied to this fibre AOM also for generating the two sidebands in the MS gate. In the current experimental path, light propagates along approximately nine meters of fibre before and inside the cryostat on which acoustic noise is not actively cancelled; vibrations along this length of fibre contribute to the coherence decay presented below (Extended Data Fig.~\ref{fig:ramsey}). 

Though the photonics are functional at room temperature and the ion trap could be operated in ultra high vacuum without cryogenic cooling, we perform ion experiments in a cryogenic environment because of the ease of achieving low background pressure via cryopumping, the possibility for  rapid trap installation/replacement (${\sim}2$ days), elimination of the need for baking of in-vacuum components, and potentially reduced electric-field noise near surfaces. The cryogenic apparatus used for ion trap experiments is similar to that described previously \cite{leupold2015bang}. Ions are loaded from neutral atom flux from a resistively heated oven within the 4K chamber, followed by two-step photoionization using beams at 423 nm and 389 nm \cite{lucas2004isotope}. The fibre feedthrough consists simply of bare 250 $\mu$m-diameter acrylate-coated fibres epoxied (Stycast 2850FT) into a hole drilled in a CF blank flange mounted on the outer vacuum chamber. In the experiments presented in this paper only the 729 nm beam for qubit control was delivered via integrated optics. Though waveguide structures for 854 nm/866 nm light were included on chip, these beams were routed together with the shorter wavelengths and delivered along conventional free-space beam paths for convenience given the previously existing optical setup. 

DC voltage sets for axial confinement, stray-field compensation, and radial mode rotation  were calculated using methods similar to those described in \cite{allcock2010implementation}, based on 3D field simulations including the effect of exposed dielectric in our design. 

We used these voltage sets to compensate stray fields in 3D by minimizing first and second micromotion sidebands of ion motion using two separate 729 nm beampaths. Light emitted by the grating has a wavevector $\vec{k_{g} }= k_0(\mathbf{\hat{z}} \cos{\theta_g}  +  \mathbf{\hat{x}}\sin{\theta_g} )$, where $\theta_g = 36^\circ$ is the emission angle from vertical and $k_0 = 2\pi/\lambda$ is the free-space wavenumber; this beam  hence allows detection of micromotion along $\hat{x}$ and $\hat{z}$. To compensate with sensitivity along $\hat{y}$, we used a second beam propagating in free-space along $\vec{k_{f} }= k_0(\mathbf{\hat{x}} \cos{\theta_f}  +  \mathbf{\hat{y}}\sin{\theta_f} )$ with $\theta_f = 45^\circ$. The calculated compensation voltage sets were used to minimize micromotion sidebands on both beam paths, which allowed us to achieve first micromotion sideband to carrier Rabi frequency ratios of $\Omega_\mathrm{MM}/\Omega_\mathrm{car} \approx 0.01$. Compensation fields applied are of order 300 V/m along $y$ and 1500 V/m along $z$; day-to-day drifts are of order 1\%. We note the requirement for beams along two different directions for compensation arises from the typical assumption that the optical field's gradient (or, for quadrupole transitions, curvature) is nonzero only along its propagation direction; this does not hold for transverse focusing to wavelength-scale spots, in which ions undergoing micromotion perpendicular to the beam propagation direction will experience amplitude modulation. Beams focused to wavelength-scale spots may thus allow sensitive compensation along all dimensions using a single beam. 

The silicon partially exposed to the ion due to openings in the ground plane near the gratings was a potential concern, as previous work has seen that unshielded semiconductor can be problematic for stable trap operation \cite{mehta2014ion}; we note that the substrate was not highly conductive or grounded in that previous work, and the grounding implemented here may be a significant difference. We  have observed charging effects that appear to be related to carrier dynamics in the silicon; in particular, we observe jumps in the motional frequency of order 10 kHz after Doppler cooling and state preparation with the 397 nm beam, which relax on millisecond timescales. These jumps were eliminated when a few mW of IR light were input into any of the waveguide channels on the chip. We attribute this to photoexcited carriers in the silicon from light scattered out of the waveguides and gratings, which diffuse through the substrate, increasing its conductivity and more effectively shorting it to ground, thereby suppressing stray fields originating from the substrate. The gate data taken in this paper were obtained with 4 mW of CW light at $\lambda = 785$ nm combined with the 729 nm pulses in the fibre coupled to input 3. It is possible that the use of heavily doped silicon would attenuate possible stray fields from the silicon; in a CMOS context, highly-doped and contacted layers at the exposed surfaces could potentially be used for the same purpose. 

Additionally, we observed kHz-level drifts in motional frequencies which relaxed on longer timescales of many minutes, which we attribute to charges trapped in or on the surface of the dielectric near the grating windows -- this charging was clearly correlated, for example, to the 729 nm beams used for the MS gates being turned on. To minimize drifts in the motional frequency, we pulsed on the 729 nm beam during the 1 ms-long Doppler cooling pulse applied in each experimental shot, which together with the 785 nm light sent to the same coupler led to a more constant optical flux through the opening and reduced these drifts to the level of a few 100 Hz. We additionally recalibrate the motional frequency every 15 s during MS gate experiments to minimize the effect of these remaining drifts. Our experiments serve as a baseline for performance achievable with no shielding of the exposed dielectric; the extent to which conductive shielding of this exposed dielectric (as recently demonstrated \cite{niffenegger2020integrated}) reduces the magnitude of these effects will be an interesting question for future experiments. 

\subsection{Contributions to Bell-state infidelity and routes to improvement}
Here we detail the error sources contributing to the Bell state infidelity achieved via the integrated MS gate (Table~\ref{tab:errors}), and discuss routes to improvement. 

The heating rate of the two-ion stretch mode at 2.2 MHz used for the MS gates was measured to be $\dot{\bar{n}} = 60(30)$ quanta/s, via fits to sideband flopping following variable wait times. For the single-loop gates implemented here, this contributes an error $\epsilon_h = \dot{\bar{n}}\tau_g/2$,\cite{ballance2017high} resulting in our estimate of $2(1) \times 10^{-3}$. We measure a heating rate for a single-ion axial mode at 1.2 MHz of ${\sim}3000$ quanta/s, indicating $E$-field noise in our device ${\sim}100\times$ higher than in other cryogenic surface-electrode traps with similar ion-electrode distance \cite{sedlacek2018distance}. 

During gate experiments, we observed an average drift in motional frequency of 200 Hz magnitude between each recalibration. Assuming a linear drift in frequency between these calibration points, we generate a probability density function for the motional frequency error during each experimental shot, which together with simulated infidelity resulting from an error in gate detuning $\delta$ results in an expectation value for infidelity of $1\times10^{-3}$.

\begin{figure}[t]
\centerline{\includegraphics[width=0.5\textwidth]{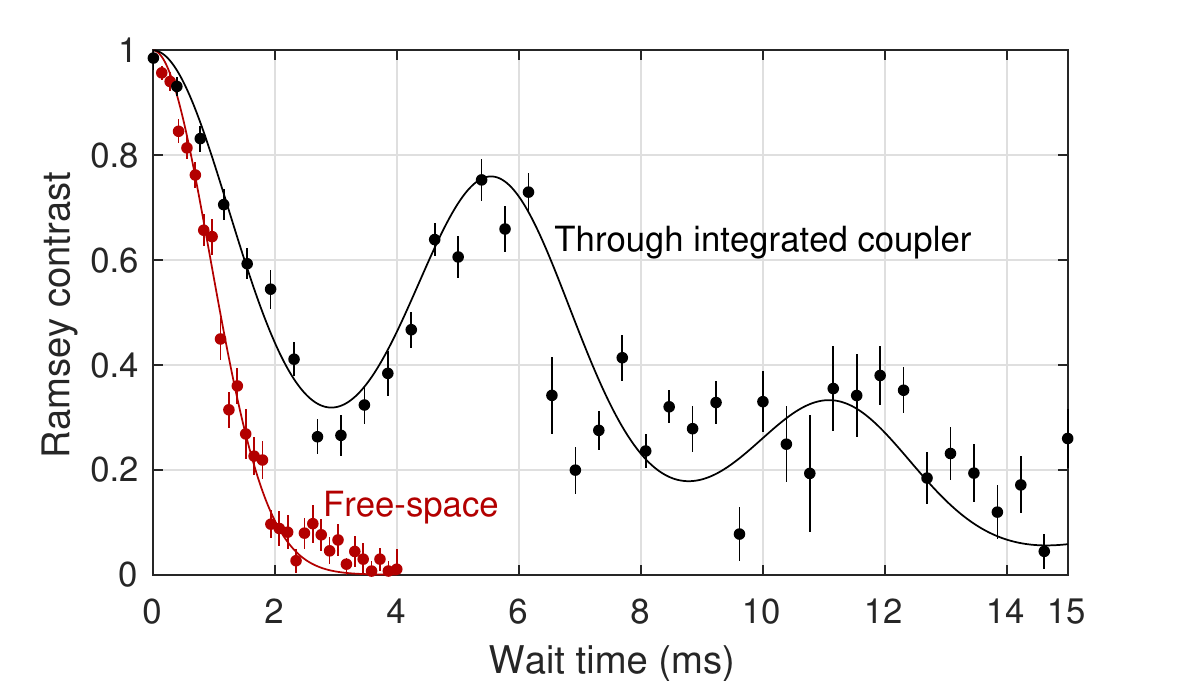}}
\vspace{0 cm}
\caption{\label{fig:ramsey} \textbf{Ramsey coherence measurements.} We apply two $\pi/2$ pulses  separated by a variable wait time, and the fringe contrast upon scanning the phase of the second pulse relative to the first is plotted to assess $T_2^*$. Data is shown using the same light guided through the in-cryostat fibres and integrated couplers (black points and fit) or through free-space (red points and fit). The fit to the data observed with the integrated coupler was used to infer laser noise parameters relevant to gate infidelity calculation; the observation of significantly faster decoherence when driving with the free-space beam (red points/fit) using the same 729 nm source indicates the integrated beam path's advantage in insensitivity to cryostat vibrations. Error bars on points represent 68\% confidence intervals on fit contrasts. }
\end{figure}

Spin coherence was assessed by means of Ramsey measurements on single qubits. Ramsey decays (Extended Data Fig.~\ref{fig:ramsey}) observed when driving the $\ket{S_\frac12,m_J = -\frac12}$ to $\ket{D_\frac52,m_J = -\frac12}$ qubit transition through the integrated gratings were fit by a model with a discrete noise component \cite{kotler2013nonlinear} corresponding to oscillations in apparent carrier frequency occurring with $175$ Hz periodicity, with a carrier frequency excursion amplitude of $2\pi\times160$ Hz, together with a slower Gaussian decay with $1/e$ time of 11 ms. We found similar decays on the $\ket{S_\frac12,m_J = -\frac12}$ to $\ket{D_\frac52,m_J = -\frac32}$ transition, which has $2\times$ higher magnetic field sensitivity, suggesting that laser frequency fluctuations (and not magnetic field drifts) are the dominant contribution. We estimate the infidelity resulting from the discrete noise component that dominates the Ramsey decay as follows. We perform numerical simulation to find the MS gate infidelity resulting from an offset carrier frequency, and average over the probability density function describing the offset during each shot for sinusoidal noise with the amplitude inferred from the Ramsey data. This results in an expectation value for infidelity of $1\times10^{-3}$, which serves as a lower bound for error as we have considered only the dominant noise component. We note that errors from drifts in both motional and carrier frequencies, together accounting for about $2\times10^{-3}$ of our Bell-State infidelity, can add coherently in sequences of multiple gates. 

The use of the stretch mode is advantageous with respect to heating, but introduces sensitivity to radial mode temperatures; variance in occupancy of the ``rocking" radial modes results in variance in the stretch mode frequency through a Kerr-type interaction \cite{roos2008nonlinear, nie2009theory}. In our experiments, for technical reasons owing to beam directions required for EIT cooling, radial modes are not ground-state cooled, and after Doppler cooling we observe occupancies corresponding to $\bar{n} \sim 12-13$ for the modes at 3.5 MHz and $\bar{n} \sim 5$ for those at 5.5 MHz; these result in an estimate of the resulting infidelity of $4 \times 10^{-4}$. 

The warm radial modes also contribute to shot-to-shot carrier Rabi frequency fluctuations, since the grating beam couples to all radial modes (these dominate the decay observed in Rabi oscillations in Fig.~\ref{fig:2ion}a and Extended Data Fig.~\ref{fig:crosstalk}a). From the same occupancies given above, we estimate an infidelity of $3 \times 10^{-4}$. 

Qubit state preparation is based on frequency-selected optical pumping on the quadrupole transition (repeatedly driving a 729 nm $\pi$-pulse starting from $\ket{S_\frac12,m_J = +\frac12}$ and repumping with the 854 nm beam to the fast-decaying $P_\frac32$ levels); we measure preparation of qubits in the starting $\ket{S_\frac12,m_J = -\frac12}$ state with infidelities $<10^{-4}$. Qubit state detection is based on applying a resonant 397 nm pulse (driving ${S_\frac12} \leftrightarrow {P_\frac12}$ transitions) for $250$ $\mu$s after each shot and thresholding the number of photon counts detected by our photomultiplier tube (PMT), to infer whether the detection event corresponded to 0,1, or 2 ions bright. The 1-bright-ion histogram is well separated from the dark distribution and contributes negligible error; however, with the optimal threshold, 0.09\% of 1-bright-ion events are mistaken for 2 bright, and vice versa, contributing $0.9\times10^{-4}$ to our Bell-State infidelity. 

The gate pulse is smoothly ramped on and off over ${\sim}5$ $\mu$s, which reduces infidelity from off-resonant excitation of the direct drive term in the MS interaction Hamiltonian \cite{sorensen2000entanglement} to negligible levels for these gate times.

\begin{figure}[b]
\centerline{\includegraphics[width=0.5\textwidth]{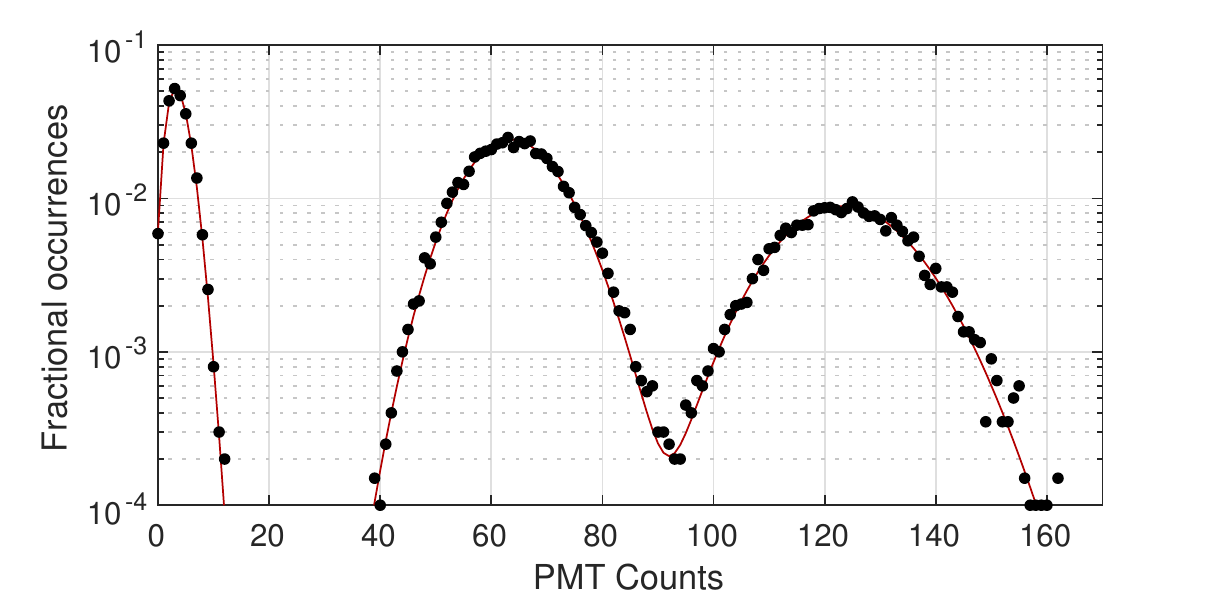}}
\vspace{0 cm}
\caption{\label{fig:histograms} \textbf{Readout histograms.} Histogram of PMT counts observed in detection events over all points in the parity scan of Fig.~\ref{fig:gate}, fitted to a sum of 3 Poissonian distributions. Each distribution corresponds to counts obtained during a 250 $\mu$s detection period from events with either 0, 1, or 2 ions in the bright state.}
\end{figure}

The effects summarized above together account for about 0.6\% infidelity. The gate fidelity could clearly be increased by means of certain technical improvements; error from motional mode can be suppressed by two orders of magnitude if improved trap materials or technical noise sources can allow heating rates comparable to the lowest-noise cryogenic surface traps \cite{sedlacek2018distance}. Shielding of the exposed dielectric e.g. with a transparent conductor \cite{niffenegger2020integrated} or perhaps with thin, semitransparent metal films should reduce motional frequency drifts, and optimal strategies for this are likely to be essential especially for trap chips delivering significant powers in the blue/UV as well. Laser noise could be reduced using improved implementation of acoustic noise cancellation on optical fibre in our system \cite{ma1994delivering}, together with technical improvements to our laser frequency stabilization. Ground-state cooling of radial modes can suppress Kerr cross-coupling and Rabi frequency fluctuations from radial mode occupancies, both of which which contribute error roughly quadratic in $\bar{n}$, by $>100\times$. Couplers and beam/field geometries in future chips for EIT ground-state cooling of all modes \cite{lechner2016electromagnetically} will be an enabling development. Aside from these possible technical improvements, multi-loop and composite-pulse gate implementations can reduce infidelity from motional heating and frequency drifts, as well as laser frequency drifts \cite{milne2020phase}. Our current experiments show no obvious limit to fidelity arising from optical integration, and it appears such implementations may assist in approaching the ${\sim}10^{-5}$ limit imposed by spontaneous emission for typical gate times using this qubit. 

\subsection{Crosstalk between trap zones}
Optical radiation brings advantages in the possibility for tight beam focuses addressing individual ions or ensembles within a larger systems. 

\begin{figure}[b]
\centerline{\includegraphics[width=0.5\textwidth]{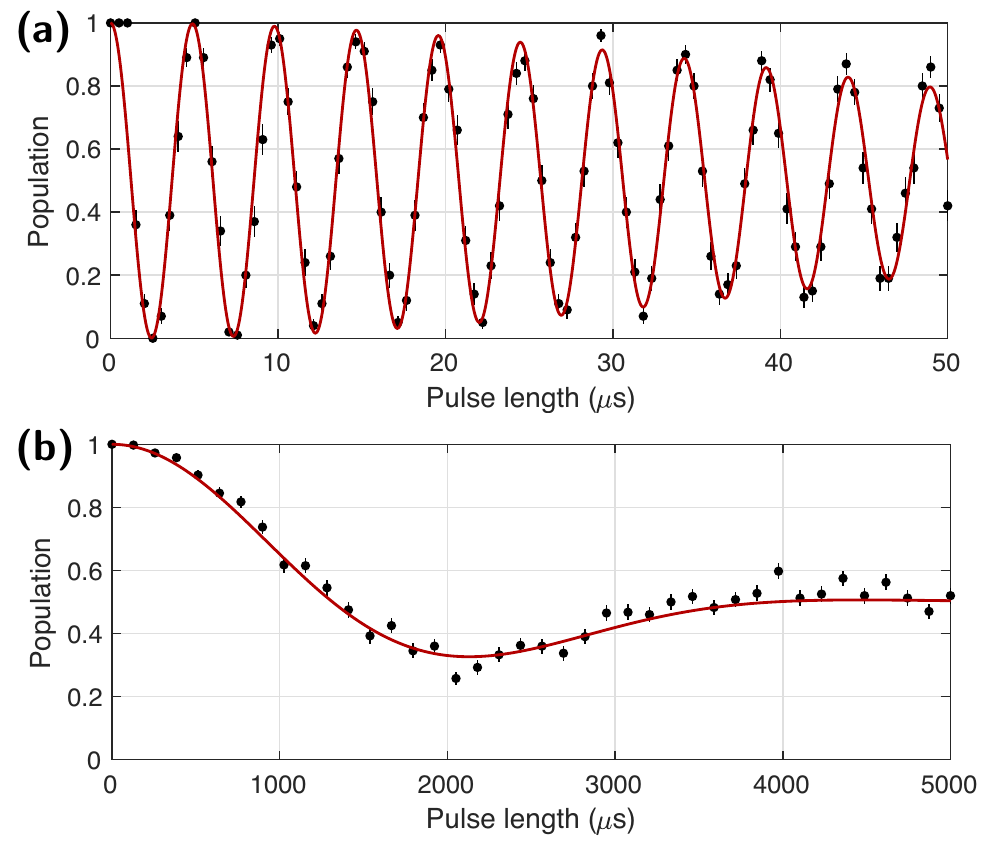}}
\vspace{0 cm}
\caption{\label{fig:crosstalk} \textbf{Crosstalk characterization.} Rabi oscillations at zone 3 (a) with light coupled to the port directly addressing this zone (input 3), and (b) with light coupled to the port intended to address zone 2 (input 5). Fits to Rabi oscillations with a Gaussian envelope decay indicate $\pi$-times of 2.4 $\mu$s (a) and 2.6 ms (b). }
\end{figure}

We quantified crosstalk levels expected in parallel operation of multiple zones in the present device by inputting light to adjacent fibre couplers and measuring the effect on the ion in the experimental trap zone 3 (zones and waveguide inputs labeled in Extended Data Fig.~\ref{fig:wafer}c). Extended Data Fig.~\ref{fig:crosstalk}a shows Rabi flopping observed on a single ion trapped in zone 3 with light coupled to input 3 (which directly feeds the grating addressing this zone). Light intended to address zone 2 would instead be coupled to input 5; when we send the same optical power to input 5, we observe Rabi oscillations with a $1000\times$ lower Rabi frequency on the ion in zone 3 (Extended Data Fig.~\ref{fig:crosstalk}b), indicating a relative intensity in this non-addressed zone of -60 dB.  

This level of crosstalk appears to arise from a combination of coupling between the waveguides on-chip (i.e. coupling from the waveguide of input 5 to input 3, and then emission by the grating in zone 3), and weak sidelobes of emission from the zone 2 grating to the ion in zone 3. Similar levels of crosstalk at -60 dB are observable when feeding in light to input 1, which propagates in a loop around the device and to no grating. The beam profile emitted in this case is probed by measuring ion response at different ion positions along the trap axis and is observed to be consistent with that emitted by the 729 nm coupler in zone 3; hence we attribute this crosstalk to coupling between the waveguides fed by input 1 and 3. Furthermore, when polarization is adjusted to minimize transmission to the loop output, the crosstalk drops by 20 dB, suggesting that the bulk of crosstalk into the zone 3 coupler comes from inter-waveguide coupling rather than scatter at the fibre-chip interface. We expect lower-loss waveguides with less sidewall scattering to reduce this effect. Crosstalk from the emission of the grating itself is particularly pronounced along the axial direction we probe here \cite{mehta2017precise}, and we expect optimized arrangements and designs to reduce this as well. A variety of composite pulse schemes can reduce the impact of coherent errors resulting from such crosstalk. 

\subsection{Hybrid Zeeman/optical qubit encoding}
Sequences of gates conducted using the optical qubit addressed in this work would require phase stability of the driving 729 nm light over the entire algorithm, a challenging technical requirement for algorithms extending beyond many seconds. The 1.1 s spontaneous emission lifetime also limits memory time achievable with this transition. 

We note that both issues may be ameliorated by hybrid approaches where, for example in $^{40}$Ca$^+$, qubits can be stored in the two $S_\frac12$ Zeeman sublevels (and single-qubit gates implemented via RF magnetic fields), with excitation from one sublevel to the $D_\frac52$ state only to implement two-qubit interactions. 

For each ion, denote the three levels involved $\ket{{0}} = \ket{S_\frac12, m_J = -\frac{1}{2}}$, $\ket{{1}} = \ket{S_\frac12, m_J = +\frac{1}{2}}$, and $\ket{{2}}$ a sublevel of the $D_\frac52$ manifold; $\ket{{0}}$ and $\ket{{1}}$ represent the long-term ``memory" qubit. $\ket{{1}}$ is mapped to $\ket{{2}}$ for optical multi-qubit gates via rotations $\hat{R}_{12}(\pi, \phi) = \exp \left(-i \pi \hat{\sigma}_{12}(\phi)/2\right)$, with $\hat{\sigma}_{12}(\phi) = \hat\sigma_{x12} \cos\phi + \hat\sigma_{y12} \sin\phi$, where $\hat\sigma_{x12}$ and $\hat\sigma_{y12}$ are Pauli matrices for states $\ket 1$ and $\ket 2$. The MS interaction $\hat{U}^\mathrm{MS}_{02}(\phi)$ between $\ket 0$ and $\ket 2$ is expressed exactly as in the main text, with Pauli matrices appropriate to these two levels; defining the global $\pi$-rotation on qubits $a$ and $b$ as $\hat{\mathcal R}_{12}(\phi) = \hat{R}_{12}(\pi, \phi)_a \otimes \hat{R}_{12}(\pi, \phi)_b$, we find the total unitary $\hat{\mathcal R}_{12}(\phi) \hat{U}^\mathrm{MS}_{02}(\phi) \hat{\mathcal R}_{12}(\phi)$ is independent of the constant laser phase offset $\phi$. In fact we note this is not unique to the MS gate and would apply to any optically-implemented unitary. 

Such an approach may allow systems to exploit the relatively low power requirements and high addressability of optical interactions, while benefiting from the long memory times and relaxed laser phase stability requirements of microwave qubits.

\subsection*{Acknowledgements}

We thank Denys Marchenko, Douwe Geuzebroek, and Arne Leinse at LioniX International for fabrication of the devices and for discussions during design. We are grateful to Vlad Negnevitsky and Matteo Marinelli for their work on the experimental control system and software used for these experiments, and Stefanie Miller for assistance in characterization of fabricated photonic devices. We thank Frank G{\"u}rkaynak at ETH for support with CAD software, the ETH FIRST cleanroom staff, and Esther Schlatter for helpful advice on epoxies. 

We acknowledge funding from the Swiss National Fund grant no. 200020165555, NCCR QSIT, ETH Z{\"u}rich, the EU Quantum Flagship, and an ETH Postdoctoral Fellowship. 

\subsection*{Author contributions}
KKM conceived the work, and designed, characterized, and assembled the trap devices. KKM, CZ, and MM performed the trapped-ion experiments in an apparatus with significant contributions from CZ, MM, TLN, and MS. KKM analyzed the data. JPH supervised the work, and KKM wrote the manuscript with input from all authors. 

\subsection*{Competing interests}
The authors declare no competing financial interests. 

\subsection*{Data availability}
The raw data generated during this study, together with analysis code employed, are available from the corresponding author on reasonable request.

\subsection*{Additional information}
$^*$Correspondences and requests for materials should be addressed to mehtak@phys.ethz.ch.

\end{document}